\documentclass[12pt]{iopart}

\usepackage{epsfig}

\newcommand{\beq}{\begin{equation}}
\newcommand{\eeq}{\end{equation}}
\newcommand{\be}{\begin{eqnarray}}
\newcommand{\ee}{\end{eqnarray}}
\def\d{\partial}
\def\+{\dagger}

\def\la{\langle}
\def\ra{\rangle}
\begin{document}
\title[Dark Matter as  Color Superconductor ]
{``Nonbaryonic" Dark Matter as Baryonic Color Superconductor }
\author{Ariel R. Zhitnitsky}
\address{Department of Physics
 and Astronomy, University of British Columbia,
  Vancouver, BC V6T 1Z1, Canada}

\begin{abstract}
 We discuss a novel cold dark matter candidate which
 is formed from the ordinary quarks
during the QCD phase transition when  the  axion domain wall
undergoes an unchecked collapse
due to the tension in the wall. If a large number of quarks
is trapped inside the bulk of a  closed  axion domain wall, the collapse 
 stops due to the internal  Fermi pressure. In this case
the system in the bulk, may reach the critical 
density when it undergoes a phase transition to a color  superconducting phase
with the ground state  being the  quark condensate,
 similar to the Cooper pairs in BCS theory. If this happens, the new state of matter
representing the diquark condensate with a 
large baryon number $B \sim  10^{32}$
  becomes  a stable soliton-like  configuration. Consequently, it may 
serve as a  novel cold dark matter candidate. 

\end{abstract}


\maketitle
\section{Introduction}
The presence of large amounts of non-luminous components in the Universe has been
known for a long time. In spite of the recent advances in the field ( see e.g. recent summary
  \cite{snowmass}), the mystery of the 
dark matter/energy remains: we still do not know what is it. The main goal of this work
is to argue that the dark matter could be  nothing but well-known quarks
which however are not in the ``normal" hadronic phase, but rather
in some ``exotic", the so-called color superconducting (CS) phase. 

This is a novel phase in QCD
when light quarks form the condensate in diquark channels, and it  is  analogous to Cooper pairs 
of electrons in ordinary superconductors described by BCS theory. There existence 
of CS phase in QCD represents our first crucial
element for our scenario to work. The study of CS phase received a lot of attention   
last few years, see original papers\cite{cs_o},\cite{cs_n}
 and recent reviews\cite{cs_r} on the subject.
 It turns out that CS phase is realized when quarks are squeezed to the density which 
is  few times
nuclear density. It has been known that this regime may be realized in nature in neutron stars
interiors and in the violent events associated with collapse of massive stars or collisions 
of neutron stars, so it is important for astrophysics. The goal of this work is to argue
that such conditions may occur in early universe during the QCD phase transition.
 Therefore, it might be important for cosmology as well.

 The force which squeezes quarks in neutron stars is gravity; the force which does
a similar job in early universe  during the QCD phase transition is 
a violent collapse  of a bubble formed from  the axion domain wall.
 If  number of quarks trapped inside of the bubble (in the bulk)
 is sufficiently large, the collapse 
 stops due to the internal Fermi pressure. In this case the system
in the bulk  may reach the critical 
density when it undergoes a phase transition to CS phase 
with the ground state  being the  diquark condensate. 
These configurations with large number of quarks   in color superconducting phase,
will  be named the QCD balls.
Therefore, an existence of the axion domain wall
represents our second crucial element for our scenario to work.
We should note at this point that the axion field was introduced into the theory to explain 
the lack
of CP violation in the strong interactions. Later on the axion field
became one of the favorite candidates for the 
cold dark matter, see original papers \cite{PQ}-\cite{DFSZ}
 and recent reviews \cite{axion_r} on the subject. 
In the present scenario the axion field plays the role of squeezer rather than 
dark matter itself. In principle, it can be replaced by some other, yet unknown fields with 
similar properties. However,  to be more concrete in estimates below we shall use 
the specific properties of   the axion field with known constraints on its coupling  constant.

We do not address the problem of formation of QCD-ball  in this letter. 
Instead we concentrate  on
the problem of stability of these objects. As we will show, once
such a configuration is formed, it will be extremely stable soliton like particle.
The source of the stability of the QCD-balls is related to the fact that
its mass $M_B$ 
becomes smaller than the mass of a collection of free separated nucleons with the same
baryon charge. The region of the absolute stability of the QCD-balls is determined by inequality
$m_N > M_{B}-M_{B-1}$ which is satisfied in some region of $B$, i.e. $B_{min}< B < B_{max}$.
The lower limit $B_{min}$ in this region determined  
 by inequality $m_N > M_{B}-M_{B-1}$ when the system becomes unstable with respect
 to decay to the nucleons. 
The upper limit $B_{max}$   is determined by the region of  applicability of our approach
when the baryon density   in the bulk becomes close to 
   the nuclear density, and therefore,  
 our calculation scheme (based
 on description in  terms of quarks )  becomes unjustified at this point.
Different  approaches   (based on consideration of hadronic rather than quark degrees of 
freedom) have to be used in this regime. 
 It could  happen that some  metastable (or even stable) states may exist
 in this low-density regime.  
 However, the  corresponding analysis   is beyond the scope
of the present work and it shall not be considered here.

Therefore, if sufficiently large number of quarks (determined mainly
by the axion properties)  is trapped inside the 
axion bubble during its shrinking, it may result in formation of an absolutely stable 
 QCD-ball with the ground state being a diquark condensate.  
 Such QCD-balls, therefore,  may serve as the cold dark matter candidate which amounts
about 30\% of the total matter/energy of the Universe, $\Omega_{DM}\simeq 0.3 $\cite{snowmass}.

 Strictly speaking, the QCD-balls being the baryonic configurations, would behave like 
nonbaryonic dark matter. In particular, QCD-balls, in spite of their QCD origin, would not 
contribute
to  $\Omega_{B}h^2\simeq 0.02$ in nucleosynthesis calculations because 
the QCD-balls would complete the formation by the time 
when temperature reaches the relevant for nucleosynthesis
region   $T\sim 1 MeV$. Once QCD-balls are formed, their
 baryon charge  is accumulated in form of the diquark condensate, rather than in form
 of free baryons, and in such a form the baryon charge is not available for  nucleosynthesis.
Therefore, the observed relation $\Omega_B\sim\Omega_{DM}$ within an order of magnitude 
finds its natural explanation in this scenario: both contributions to $\Omega$ originated from  the 
same physics at the same instant during the QCD phase transition. As is known, this fact is extremely
difficult to explain in models that invoke a dark matter candidate not related to baryons.

Before we continue the description of our proposal we would like to make few 
  comments on what have happened on the theoretical side during the last few years, 
which  are  crucial elements in our present discussions, and which were  not available
 to researchers earlier. 

First of all, there existence of the axion domain
walls, related to the symmetry under
discrete rotations of the so-called  $\theta$ angle $\theta\rightarrow\theta+2\pi n$
has been known for a long time since \cite{Sikivie}.
  However, the structure of the domain wall considered
in  \cite{Sikivie} had only one typical scale,  $m_a^{-1} \gg 1 fermi$. 
Therefore, the quarks, even if they were trapped inside the bubble at the very first moment,
 could easily penetrate through such domain wall configuration during the bubble evolution. 
In this case the axion domain wall ( without support of
the fermi pressure from the bulk) would completely collapse.
What was realized only quite recently, is the fact that the axion domain walls have 
actually sandwich substructure
on the QCD scale $\Lambda_{QCD}^{-1}\simeq 1 fermi$. Therefore, the fermions
which are trapped inside the bubble at the very first instant,
can not  easily penetrate through  the  domain wall due to this  QCD scale substructure,
and will likely stay in the bulk, inside the bubble. In this case, 
the collapse of the axion domain wall stops due to the fermi pressure in the bulk.
The arguments ( regarding there existence of the
QCD scale substructure inside the axion domain walls) are based on analysis\cite{FZ}
of QCD in the large $N_c$ limit with inclusion of the  $\eta'$ field\footnote{
Uniqueness of the $\eta'$ field in this problem is related to the special
structure of interaction of  the axion field $\theta(x)$ and the singlet  $\eta' (x)$  field 
in low energy description of QCD when only a special combination $[\theta(x)- \eta'(x)]$  
is allowed to enter the low energy QCD Lagrangian. } and  independent
analysis \cite{SG} of supersymmetric models
where a similar $\theta$ vacuum structure occurs.
 
The second important element  of our proposal not available earlier,
is related to the recent advances\cite{cs_n},\cite{cs_r} in understanding of CS phase.
The fact that the color superconducting phase may exist at high baryon density
was discussed a while ago\cite{cs_o}, however it was not a  widely accepted phenomenon
 until recent papers\cite{cs_n} where a relatively large superconducting gap 
$\Delta\sim 100 MeV$ with a large critical temperature $T_c\simeq 0.6 \Delta$ were advocated.

To conclude the Introduction we should remark here that the idea that some 
quark matter, such as strange quark ``nuggets" may play a role of the dark matter,
was suggested long ago\cite{Witten}, see also original papers\cite{Jaffe}  
and relatively recent review\cite{Madsen}
 on the subject. 
The idea that soliton-like
configurations may serve as a dark matter, is also not a new idea\cite{Lee}. 
Most noticeable example is being
Q-balls\cite{qball}. The idea that the dark matter may be just  solitons containing large 
 baryon  (or even antibaryon) charge is, again, an old idea\cite{Widrow}, see also
\cite{baryogenesis}. The new element of this proposal
 is the observation that one can  accommodate all the nice
properties (discussed previously \cite{qball}- \cite{baryogenesis}) without invoking 
any new fields and particles (apart from the axion).
Rather, our QCD-balls formed from the ordinary quarks
which however are not in the ``normal" hadronic phase, but rather
in  color superconducting  phase when  squeezed quarks organize a single coherent state 
described by the diquark Bose--condensate, similar to the Cooper pair condensate
in BCS theory in conventional superconductors.

In many respects ( in terms of phenomenology)
the QCD balls are similar to strangelets\cite{Witten}-\cite{Madsen}
 with few important   differences, see below:\\
$\bullet 1.$ In our proposal
the first order QCD phase transition is not required 
for the formation of  the QCD-balls. Axion domain walls of a large size (in comparison with
a typical QCD scale) are able to  form  the large bubbles. These bubbles,
 filled by $u, d, s$ quarks, play the same role as the bubbles formed during
the first order phase transition as discussed in\cite{Witten}.\\
$\bullet 2.$   The Stability of strange quark matter at  zero external pressure, as described 
in\cite{Witten}-\cite{Madsen}, is highly  model dependent result. In particular,
the stability of strangelets is very sensitive  to the
magnitude of the bag constant within MIT bag model calculations.
The idea  which is advocated in the present work has a new element, the external pressure
due to the axion domain walls. With  this new element  the  stability 
of the system is very likely to occur in very wide region of the parametric space
even in the models which would not support strangelets in the absence of the external pressure.\\
$\bullet 3.$ The bulk of the QCD ball is in the superconducting phase. This property obviously influences
the phenomenology of how the  QCD balls interact with a normal matter. In  particular,
if the energy of a hadron which hits the QCD ball is smaller than the 
superconducting gap $\Delta$, the hadron can not penetrate into the bulk and excite
the internal degrees of freedom of the system, but
rather it will be reflected (the so-called, Andreev reflection). Similar property
is also true for the strangelets\cite{Witten}. 
The elastic cross -section of hadrons on the QCD balls is large, of the order
of the geometrical size of QCD balls; the inelastic cross -section 
(when internal degrees of freedom are excited) is almost identically zero
for small energies as mentioned above.
Electromagnetic interactions of photons
with QCD balls contain the standard fine structure constant $\alpha$ with an  addition
  suppression due to the neutrality of the CFL dense quark matter.
Such    features for the interactions imply that if
the QCD ball with small velocity $v/c\sim 10^{-3}$ enters the Earth, it will not decay by
exploding.  Rather
it will   go through the Earth and exit on the opposite side
of the Earth leaving behind the shock waves. 
It is tempting to interpret the recent  
 seismic event with epilinear source\cite{Teplitz}  as the process which involves
 the  dark matter particle, similar to the QCD ball.\\
$\bullet 4.$ There is a maximum size
 of the QCD-ball above which such an object can not be formed and can not be absolutely stable.
This is due to the fact that for very large system the axion domain wall pressure
becomes a negligible factor  which can not stabilize the system.\\
 $\bullet 5.$The property on a maximum size mentioned
 above has a profound
phenomenological consequence. Indeed, 
if one assumes that   stable state as described in\cite{Witten}  exists,
 the strangelets can collide with an   ordinary neutron star which results in formation of a quark
star.  In such a case all neutron starts would be transformed into quark stars.
In our proposal, when the maximal 
size of the QCD ball is determined by the external axion domain wall
pressure, this transition (from neutron stars to quark stars) does not  happen as a routine effect. \\
$\bullet 6.$  If the size of the QCD ball slightly exceeds the maximum critical size,
it becomes metastable, rather than stable configuration. Such QCD balls could also be 
interesting particles
 for the dark matter phenomenology, see footnote on page 14  and discussions in the Conclusion.

\section{QCD-balls} 
Crucial for our scenario is the existence of a squeezer, axion domain wall which will be formed
 during the QCD phase transition. As is known, there are many types of the axion domain walls, 
depending on a model. We assume that the standard problem of the 
domain wall dominance is resolved in  some way as discussed previously
in the literature, see e.g.\cite{axion_r},\cite{Sikivie1},
and we do not address this problem in the present paper\footnote{It is widely accepted that
the domain walls in the so-called, N=1 axion model will be eaten up by the axion strings at
a very high rate. That is true for the axion walls bounded by strings.
 However, if a  domain wall is formed as a closed surface, the probability for such a wall to decay 
 is extremely small. Therefore, such domain walls   
 in $ N=1 $ model can play the same role in our scenario as stable domain walls in $N\neq 1$ models.
 Besides that,
 $ N=1 $ model  has a nice property that the domain wall dominance problem is automatically resolved.}.
We also assume that the probability
of formation of a closed bubble made from the axion domain wall
is non-zero\footnote{We do not attempt to develop 
a quantitative theory of the formation 
of the QCD-balls in this work; It is sufficient for our following discussions  
 that this probability is finite.}. 
 We also assume that quarks which are trapped in the bulk, can not
 easily escape  
 the interior when the bubble  is shrinking. 
 In different words, the axion domain wall is not 
transparent due to the QCD sandwich structure of the wall  as 
 discussed in \cite{FZ},\cite{SG}. 
 The collapse is halted due to the Fermi pressure.
 Therefore, we assume that  a large number of quarks remains in the bulk,  inside  the bubble
 when the system reaches the equilibrium. 
\subsection{Equilibrium}
The equilibrium is reached  when the Fermi pressure equals 
 the surface tension and pressure due to the  bag constant $E_B$.
To put this condition on the quantitative level, we 
represent the total energy $E$ of a QCD-ball with the fixed baryon charge $B$, 
in the following way,
 \be
\label{1}
E = 4\pi\sigma R^2+ \frac{g\mu^4}{6\pi}  R^3  +\frac{4\pi}{3}E_B  R^3 ~~~~~~~~
\\ \nonumber
B= g V\int^{\mu}_0\frac{d^3p}{(2\pi)^3}= \frac{2g}{9\pi}\mu^3R^3,~~
\mu=\left(\frac{9\pi B}{2 g R^3}\right)^{\frac{1}{3}}, \nonumber
\ee
where we assume the quarks to be massless. We assume that the
relativistic fermi gas is  non-interacting in the first approximation, see corrections
due to the interactions below.
In this formula
 $\mu$ is the Fermi momentum of the system  to be expressed in terms of the fixed
baryon charge $B$ trapped in the bulk; $R$ is the
size of the sysytem; $g$ is the degeneracy factor, $g\simeq 2N_cN_f =18$
 for massless degrees of freedom; $E_B$ is  bag constant 
 which describes the  difference in vacuum energy
between the interior and exterior. The bag constant 
is a phenomenological way to  simulate the confinement. 
Finally, $\sigma\simeq f_a m_{\pi}f_{\pi}$ is  the axion domain wall tension
with $f_a\sim (10^{10}-10^{12} ) GeV$ being constrained by the axion search experiment. 

In  what follows, it is convenient to introduce dimensionless scaling variable $x$,
as follows, $x\sqrt[3]{B}= R \sqrt[4]{E_B}$ such that energy per quark $\epsilon_{tot}\equiv E/B$
can be expressed in the following simple way in terms
of dimensionless parameters $x$ and $\sigma_0$,
 \be
\label{2}
 \epsilon_{tot}(x)\equiv\frac{E}{B} = E_B^{1/4}\left(
 4\pi \sigma_0 x^2+ \frac{3}{4x}  \sqrt[3]{\frac{9\pi}{2g}}
  +\frac{4\pi}{3}x^3
\right) \\   \nonumber
x\equiv R\frac{E_B^{1/4}}{B^{1/3}}, ~~~~~~~~~~~
\sigma_0\equiv \frac{\sigma}{B^{1/3}E_B^{3/4}}. ~~~~~~~~~~~
\ee
The minimization of this expression $ {\d\epsilon_{tot}(x)}/{\d x}|_{x=x_0}=0$ determines 
the stability radius $x_0$ which  fixes the energy of the system
at the equilibrium, $\epsilon_{tot}(x_0)$. In particular, 
if one neglects $\sigma_0$ in eq. (\ref{2}) originated from  the axion domain wall tension, one reproduces
the well known results, $x_0\simeq 0.48, ~\epsilon (x_0)\simeq 1.9 E_B^{1/4}$. Such a relation means
that if  $E_B$ is relatively small  such that  the energy per quark is less than
$m_N/3$, the configuration becomes an absolutely stable state of matter\cite{Witten}-\cite{Madsen}.

In  eqs.(\ref{1}, \ref{2})  we have neglected many important contributions   
which  can drastically change  the  results.   We shall review the role of these contributions
 below.
 The main goal of this subsection is  the incorporation of these contributions
    into eqs.(\ref{1}, \ref{2}). 
First of all, in eq. (\ref{1}) we neglected 
the quark-quark interaction on 
the Fermi surface, which brings the system into  superconducting phase for relatively
 large  baryon density \cite{cs_n}. 
 The corresponding contribution $\Delta E_{int}$
to  the total energy (\ref{1}) is negative and at asymptoticaly large $\mu$
is equal to\cite{Krishna}, 
\be
\label{3}
\Delta E_{int}=-\frac{3\Delta^2\mu^2}{\pi^2}\cdot(\frac{4\pi}{3}R^3)
\ee
 The negative sign of $\Delta E_{int}$
 is quite obvious: the formation of the diquark condensate
due to the quark-quark interaction lowers the energy of the system.
 For appropriate treatment of this term
 one should express $\mu $ as a function of $B, R$ according to the 
relation (\ref{1}) and substitute this 
 into  eq. (\ref{2}). In principle, one should also take into account that
 the superconducting gap $\Delta(\mu)$ also strongly varies with $\mu$ (and therefore, with $R$)
 in the relevant region of  $\mu$.
  However, in what follows  we shall ignore this dependence for numerical estimates  
  and shall treat $\Delta \simeq 100 MeV$ as constant. 
Our last remark regarding eq. (\ref{3}). This formula was derived for 
very large $\mu$. Nevertheless for illustrative purposes we shall    use the
 expression for $ \Delta E_{int}$   for  small $\mu$ as well.
We shall see that 
in the relevant region of densities the contribution $\Delta E_{int}$ does not exceed
$15\%$. This    somewhat justifies the use of expression 
 (\ref{3}) for our numerical estimates which follow.
With all these reservations in mind, we account the additional contribution
to energy per quark,  describing  the quark-quark interaction on 
the Fermi surface by adding $\Delta\epsilon_{tot}^{int}$ into eq. (\ref{2}) in the following way
\be
\label{3a}
\Delta \epsilon_{tot}^{int}
 = -E_B^{1/4}\left(\sqrt[3]{\frac{4}{\pi}}\cdot\frac{\Delta^2}{\sqrt{E_B}}\cdot x\right) ,
\ee
where we expressed everything in terms of dimensionless parameter $\frac{\Delta^2}{\sqrt{E_B}}$ 
and dimensionless variable $x$.

  Now we want to consider the modification of eq. (\ref{1}) which
   is related to the  actual variation of the  bag ``constant" $E_B$ with $\mu$.
 To explain  the physical meaning of this effect, 
we remind the reader that the  bag ``constant" $E_B$  
 describes the  differences of vacuum energies in the interior and exterior regions. 
It  is a phenomenological
way to  simulate the confinement.
    The bag ``constant"     contribution goes with the positive sign to $E$, see eq.(\ref{1}).
    The physical reason for this sign is obvious: the vacuum energy outside the bubble
    is lower than inside, thus the positive  contribution to $E$, 
in contrast with the interaction term, $-\frac{3\Delta^2\mu^2}{\pi^2}$
discussed above.

   Our main point is as follows:  the contribution related to 
    $E_B$  can be expressed  formally
  in terms of the difference between the
   vacuum condensates calculated at zero (exterior) and non-zero (interior)  baryon densities.
   The most important contribution to $E_B$ is due to the gluon condensate, 
   such that $E_B (\mu)\sim \la \frac{b\alpha_s}{32\pi}G_{\mu\nu}^2\ra_{\mu=0}-
   \la \frac{b\alpha_s}{32\pi}G_{\mu\nu}^2\ra_{\mu\neq 0}$ with $b=\frac{11}{3}N_c-\frac{2}{3}N_f$
where we used the well-known expression for the conformal anomaly in QCD in the chiral limit.
   We do not know $ E_B (\mu)$ as a function of $\mu$  
   for the  relevant region of the baryon density. However we do know
   the behavior of this quantity for relatively small densities corresponding to
   the nuclear matter densities\cite{Cohen}, 
\be
\label{condensate}
\frac{\la \frac{\alpha_s}{\pi}G_{\mu\nu}^2\ra_{\mu\neq 0}}{ \la \frac{\alpha_s}{\pi}G_{\mu\nu}^2\ra_{\mu=0}}
\simeq 
  1-\frac{( 0.65 GeV) \rho_N}{  \la \frac{\alpha_s}{\pi}G_{\mu\nu}^2\ra_{\mu=0}}\simeq
1-\frac{\rho_N}{(264 MeV)^3}
\ee
 where $\rho_N $ is
   baryon density, and the magnitude for the gluon condensate is known to be,
 $ \la \frac{\alpha_s}{\pi}G_{\mu\nu}^2\ra_{\mu=0}\simeq 1.2\cdot 10^{-2} GeV^4$.
 As expected the gluon condensate (and therefore,
   the absolute value of the vacuum energy) decreases when the baryon density increase.
 Similar formulae are known for the 
   chiral quark condensate where for the small densities one can derive
   the following relation  $\frac{\la\bar{q}q\ra_{\mu\neq 0}}{\la\bar{q}q\ra_{\mu= 0}}
   =1-\frac{\sigma_N \rho_N}{m_{\pi}^2f_{\pi}^2}$ with sigma term measured
   to be $\sigma_N\simeq 45 MeV$ see \cite{Cohen} for the details.
One should emphasize here that the 
formula (\ref{condensate}) describing the 
variation of the gluon vacuum condensate at small 
baryon densities  $\rho_N $,
is a direct consequence of the QCD low energy theorems.
It is a firm result of QCD,  not based on any model dependent considerations,
and should be accepted as it is.

   More specific information on the
bag ``constant" $E_B$ contribution as function of $\mu$  
in the entire region of   of $\mu$ can be calculated
   in some non-physical models such as  QCD with two colors, $N_c=2$\cite{ssz_u}.
Such a knowledge can not be literally used for our numerical estimates which follow,
however it can  be  used for modeling the functional dependence of 
   the  vacuum energy.
  
Therefore, we want to model  two properties  discussed above
 in order to incorporate them into the corresponding  eq. (\ref{2}).
First,  the bag constant contribution must vanish
when the baryon density in the bulk vanishes. This corresponds
to the case when vacuum energy inside and outside of the bubble is the same, 
and therefore, it should be 
no additional vacuum energy contribution to the equation for the equilibrium. Secondly,
the bag constant contribution should vary with density as we discussed above.

Our first parametrization
is motivated by analysis\cite{ssz_u} of the vacuum condensates
in QCD-like theories at finite baryon density as a function of $\mu$.
If we assume  a similar behavior  in real QCD than we 
   should replace   the bag constant $E_B$
by the expression $E_B\rightarrow E_B(1-\frac{\mu_c^2}{\mu^2})$ for $\mu \geq \mu_c$
and $E_B\rightarrow 0$ for $\mu \leq \mu_c$, where $\mu_c$ would correspond to a  magnitude
of the critical
chemical potential at  which the baryon density vanishes.
In  QCD, one expects that this is to happen at
$\mu_c\simeq 330 MeV$.

   As before, one should  express the corresponding contribution
to $\epsilon_{tot}$
 in terms of fixed baryon charge  $B$ and radius $R$, such that the 
   bag`` constant" contribution actually becomes a complicated  function of
   $B, R$.  In terms of dimensional parameter $x$ 
the corresponding contribution to (\ref{2}) is accounted for by the following replacement,
\be
\label{3b}
E_B^{1/4}\frac{4\pi}{3}x^3\Rightarrow E_B^{1/4}\frac{4\pi}{3}x^3\cdot
\left(1-(\frac{4}{\pi})^{2/3}\cdot\frac{\mu_c^2}{\sqrt{E_B}}x^2\right)
\ee
Let us emphasize: we are not attempting to solve a difficult  problem of 
evaluation of nonperturbative vacuum energy as a function of $\mu$ in QCD.
Rather, we want to make some simple estimates to account for this effect in order to analyze
the stability of QCD balls later in the text.

We want to be confident that 
the results on stability of QCD balls (to be discussed later)  are not sensitive to the
 specific parameterization  (\ref{3b}) motivated by the study  of
QCD with two colors. Therefore, we would like to have a different,
 independent parameterization of the same effect to  be used in our stability analysis.
We   make use of eq.(\ref{condensate}) which is valid   for small densities $\rho_N$.
This formula gives us an idea about typical variation of vacuum condensates 
when the baryon density changes. We assume that 
the vacuum energy difference in QCD (the bag ``constant" contribution in eq. (\ref{2}))
can be expressed in terms of different vacuum condensates with the
typical scale for the variation  given by eq.(\ref{condensate}).

We want to implement the QCD property  (\ref{condensate}) into the MIT bag model. 
If the phenomenological numerical magnitude for the bag constant $E_B$
were closed to the numerical value for the vacuum energy 
$ \la \frac{b\alpha_s}{32\pi}G_{\mu\nu}^2\ra\simeq (340 MeV)^4$ we could literally use 
eq (\ref{condensate}),  such that 
  the bag constant contribution can be parameterized
as follows, $E_B(\rho_N) \simeq E_B\frac{\rho_N}{(264 MeV)^3}$.
Unfortunately, these two are very different numerically, and we will introduce
the corresponding correction factor
 $r\equiv \sqrt[4]{\la \frac{b\alpha_s}{32\pi}G_{\mu\nu}^2\ra  / E_B}\simeq (340 MeV)/(150 MeV)\simeq 2.25$
in our implementation of QCD property 
(\ref{condensate}) into the MIT bag model, see below.

 Still,  formula  $E_B(\rho_N) \sim  \rho_N $
  can not be used literally  for our purposes
 because we need an expression for the bag ``constant" contribution
which goes to constant $E_B$ at large densities,
$E_B(\rho_N) \rightarrow E_B$. A simple model which satisfies this requirement
 is to make the following replacement,
\be
\label{3d}
E_B(\rho_N) \simeq E_B\frac{r^3\rho_N}{(264 MeV)^3}
\Rightarrow\frac{E_B}{\left(1+   \frac{(264 MeV)^3}{r^{3}\rho_N}\right)},
\ee
where we introduced the correction factor $r$ to match the  scales.
   As before, one should  express  the bag ``constant" contribution proportional to
(\ref{3d})
 in terms of a  fixed baryon charge  $B$ and radius $R$.
We shall analyse the corresponding equation 
(\ref{2}) with improvements (\ref{3d}) in the next subsection. To anticipate the events,
one should mention that our two models (\ref{3b}, \ref{3d}) describing the effect 
 of the bag ``constant" variation with baryon density    
 lead to the similar results, see below.

The next approximation
we have made in eqs.(\ref{1}, \ref{2})  is related to the  assumption
of  a  thin-wall approximation for the domain wall. This 
  may not be well justified assumption because the typical width   of the domain wall
 and the size of QCD ball could be  the same order of magnitude,
 such that  thin-wall approximation 
is failed.  
However, we neglect these
complications at this initial stage of study. Nevertheless,
 we  do not expect that this effect can drastically change our qualitative results which
follow. 
  
   We also neglected  in eqs.(\ref{1}, \ref{2}) all complications   
related to the finite magnitude of the quark masses, first of all $m_s$, which 
result in additional $K$ condensation along with diquark condensation in CFL phase\cite{Thomas}. 
Finally, the expression for  the energy $E$ with corrections
 (\ref{3},\ref{3b}),   changes the simple  relation (\ref{1})
between baryon charge $B$ and chemical potential $\mu$
according to the standard thermodynamical relations, 
$B=-\frac{\partial F}{\partial\mu}$, where $F=E-\mu B$ is the free energy.
 However, we checked that these changes are 
relatively small(   do not exceed $5 \% $   
in the relevant region of $\mu$). Therefore, in what follows, 
in order to avoid the technical complications in the qualitative analysis,
 we use a simple algebraic
expression (\ref{1}) which is formally  valid only for noninteracting quarks, $B\sim\mu^3$,
but numerically remains a good approximation in a large region of $\mu$.
This allows us to use the dimensional variable $x$ which we introduced before for
the non-interacting case.
Let us repeat again: we do not attempt to solve the problem quantitatively 
with all   uncertainties in parameters discussed above;
rather, we want to give some  qualitative arguments  demonstrating   
that stability region might occur in the wide region of $B$ with 
realistic choice of parameters specified below.

With all these reservations regarding eqs.(\ref{1}, \ref{2}) in mind 
we  express the   energy of a QCD-ball per baryon charge $B$ in units
of $\sqrt[4]{E_B}$ , 
as follows
 \be
\label{4}
 y(x)_{tot}\equiv 
E_B^{-1/4}\epsilon_{tot}(x)  = 
  \frac{4\pi}{3}x^3\left(1-(\frac{4}{\pi})^{2/3}\frac{\mu_c^2}{\sqrt{E_B}}x^2\right)\\ \nonumber
+\left(
 4\pi \sigma_0 x^2+ \frac{3}{4x}  \sqrt[3]{\frac{\pi}{4}}
-\sqrt[3]{\frac{4}{\pi}}\cdot\frac{\Delta^2}{\sqrt{E_B}}\cdot x
\right). 
\ee
In this formula, in comparison with eq.(\ref{2}),
  we took into account  the effect describing  the quark-quark interaction on 
the Fermi surface given by eq. (\ref{3a}) and the effect of the variation
of the vacuum energy with baryon density, given by eq. (\ref{3b}).
 
The equilibrium condition $ \partial \epsilon_{tot} (x=x_0)/ \partial x=0$ determines the 
 radius $x_0$ of the QCD ball with   baryon charge $B$. We shall analyze this
condition in the next subsection; now we want to constraint $x_0 \leq \bar{x}$ 
to be considered. The constraint follows from the condition that the baryon density should be 
relatively large. In this case our treatment of the problem 
by using the quark degrees of freedom, eq.(\ref{4}), rather than hadronic degrees of freedom,
is justified.
The baryon number density  $\rho_N $ for the QCD ball configuration  is given
 by\footnote{Our normalization for the baryon charge corresponds to $B=1$ for the quark, 
thus factor $B/3$ in eq. (\ref{density}).}, 
\be
\label{density}
 \rho_N \equiv \frac{B}{3V}  = \frac{ E_B^{3/4}}{4\pi x^3} \gg n_0,~~~
n_0\simeq (108 MeV)^3,
\ee 
which gives upper limit $\bar{x}$ above which our approach is not
 justified.
Numerically, with our choice of parameters, see below,  $\bar{x}\simeq 0.6$,
and therefore, any solution  $x_0$ of the
 equilibrium condition $\partial \epsilon_{tot} (x=x_0)/ \partial x=0$  
must satisfy to the constraint $x_0\leq \bar{x}\simeq 0.6$. 

\subsection{Stability of QCD balls} 
  As expected,  
the equation describing the equilibrium $\partial \epsilon_{tot} (x=x_0)/ \partial x=0$  
has a nontrivial solution (minimum) in a large region of parametrical space
deterimed by parameters $E_B, \sigma, \Delta, \mu_c, B$. It is not our goal
to have a complete analysis of this allowed region of solutions.
 Rather, we shall make a specific
choice for all parameters except for the baryon number $B$ and analize the stability condition
as a function of $B$. We shall also comment on results with $\sigma = 0$
 corresponding to pure QCD configuration without any involvement of the axion field
(case considered previously in MIT bag model, \cite{Witten}- \cite{Madsen}).
The first step is to calculate 
the point $x=x_0$
 which is determined by equation
$\partial \epsilon_{tot} (x=x_0)/ \partial x=0$.   
The next step is to analyze the stability of the obtained configuration
as a function of external parameters.
Condition when the QCD-ball becomes an absolutely stable object can be 
derived from the following arguments. 
 Total energy per quark $\epsilon_{tot}(x_0)$ in eqs. (\ref{2}, \ref{4}) is a combination
of two factors: the first one, $\epsilon_{QCD}(x_0)$, is due to the strong interactions;
the second factor, $\epsilon_{axion}(x_0)$ is mainly due to the axion domain wall 
tension\footnote{the QCD contribution to $\sigma$ due to the $\eta'$ and pions 
is suppressed by a factor $f_{\pi}^2/f_a^2\ll 1$.}, i.e.
$\epsilon_{tot}(x_0)=\epsilon_{QCD}(x_0)+\epsilon_{axion}(x_0)$, with
$\epsilon_{axion}(x_0)\equiv E_B^{1/4}\left(4\pi \sigma_0 x_0^2\right)$
and $\epsilon_{QCD}(x_0)$ is determined by rest  of terms in eq. (\ref{4}).
The absolute stability of the system implies that a nucleon can not leave a system
because the energy of the configuration with baryon charge $B$
is smaller than the energy of configuration of charge $B-3$ plus energy
of a nucleon with baryon charge $B=3$ and energy of the  axion emission. 
Such a  situation is analogous to the three dimensional quantum mechanical
problem with an effective  potential being   a step-function and the energy of the bound 
state is lower than the potential energy at the large distances. In this case   
a particle obviously can not leave the system.

 We should emphasize here that the quarks
can not leave the system   due to  the energetic conditions 
which take place after the QCD ball is formed. In different words, the stability occurs 
due to the differences in properties inside/outside of the QCD ball, and 
not due to  the features of the original axion domain wall.
The axion domain wall already had
played its role during the formation period when  a large number of quarks could not
 escape the system 
and were  trapped in the bulk during the collapse of the wall.  
A similar situation when a configuration may become a  stable one due to a difference in 
conditions (inside/outside the bulk) was discussed 
long ago\cite{Lee} as an example of a non-topological soliton in
 quantum field theory. We further comment on the similarities with non-topological solitons  
later in the text.  

It is quite obvious that the axion domain wall 
with a typical correlation length $\sim m_a^{-1} \gg \Lambda_{QCD}^{-1}$
  can not produce nucleons by itself when it shrinks   due to the nucleon emission. 
Instead, typically, the axion domain wall reduces its size by emitting the  axions while 
the nucleon leaves the system.   
In this case the term 
$\epsilon_{axion}(x_0)\equiv E_B^{1/4}\left(4\pi \sigma_0 x_0^2\right)$
  is responsible for the emission of axions
rather than production of nucleons. As a result of this, this term should be ignored
for the analysis of the stability.
However, with exceedingly small probability  
the emitted axion, in principle,  can be absorbed by the  nucleon which leaves the system. 
In this case the energy, in principle, can be transformed from the axion domain wall to the
produced nucleon and the term 
$\epsilon_{axion}(x_0)\equiv E_B^{1/4}\left(4\pi \sigma_0 x_0^2\right)$
should be accounted in the energy budget  for analysis of the decay.
We estimate in appendix that the probability for the corresponding absorption of
the axion by the leaving nucleon is negligible.
Therefore in what follows we neglect this process. 

The relevant term which describes the emission of    nucleons
is the one related to the QCD  physics
i.e. $\epsilon_{QCD}(x_0)$. Therefore, the condition
when configuration becomes a
  sufficiently stable (with the life time
exceeding  the life time of the Universe,  see Appendix for details) 
 is determined from the following inequality
\be
\label{stability}
 \epsilon_{QCD}(x_0)< \frac{m_N}{3},~~ \frac{\partial \epsilon_{tot} (x)}
{\partial x}|_{x=x_0}=0, ~~ x_0 < \bar{x},
\ee
 where the last condition follows from (\ref{density}).

To analyse eq.(\ref{stability}) we shall accept the following
magnitudes for the dimensional parameters:
\be
\label{parameter}
\Delta \simeq 100 MeV;~~\sigma\simeq 1.8\cdot 10^8 GeV^3; \\ \nonumber
\mu_c\simeq 330MeV;~~
E_B\simeq (150 MeV)^4.
\ee
Having  these external parameters fixed, we left with the only one unknown 
number, the baryon charge $B$, which eneters $\sigma_0$ in our dimensionless 
parametrization (\ref{2},\ref{4}). We shall treat $\sigma_0$ as a free parameter
and our goal is to find the region of $\sigma_0$ when conditions (\ref{stability})
are satisfied.
As we discussed above, we shall use two different models to account
the effect of  the  variation of the bag
constant contribution with   density, see eqs. (\ref{3b}, \ref{3d}).

Having defined our stability condition (\ref{stability}), external 
 parameters (\ref{parameter}) and  
two simple models  accounting the effect of  the  variation of the bag
constant, eqs. (\ref{3b}, \ref{3d}), we reduce our problem to 
analysis of  dimensionless functions, $y_{QCD}^{(1)}(x)$ and 
$y_{QCD}^{(2)}(x)$ defined as follows, see eqs. (\ref{3b}, \ref{3d}, \ref{4}),
  \be
\label{y}
 y_{tot}(x)\equiv y_{QCD}^{(1,2)}(x)+y_{axion}(x);~~~
y_{axion}(x)\equiv
 4\pi \sigma_0 x^2    \\
\label{5a}
y_{QCD}^{(1)}\equiv \frac{0.69}{x} +4.2x^3\frac{1}{1+6 x^3}-{0.48}{x},~~~~~~~~~~~~~
\ee
\beq
\label{5b}
y_{QCD}^{(2)}\equiv \frac{0.69}{x} +4.2x^3(1-5.68 x^2)-{0.48}{x}, ~~~~~~~
\eeq
where three consequent terms describe: 
the fermi pressure,  the bag constant contribution accounting the
variation of the vacuum energy with the baryon density
(\ref{3d},\ref{3b}), and, finally, the quark-quark
 interaction on the fermi surface (\ref{3a}) correspondingly.
Stability condition (\ref{stability}) in dimensionless variables becomes
\be
\label{stability1}
 y_{QCD}^{(1,2)}(x_0)< \frac{m_N}{3\sqrt[4]{E_B}}\simeq 2.1,~~ \frac{\partial y_{tot} (x)}
{\partial x}|_{x=x_0}=0.
\ee
Before we  discuss some specific numerical results which follow from analysis of eqs.
(\ref{5a} -\ref{stability1}),
we would like  to list some general  model-independent properties of the solutions. 
We believe that the properties listed below are quite  common
features of the QCD balls, which
  likely to remain untouched even
in a more general treatment of the problem when many additional effects are included
(some of these effects were mentioned above).
\begin{figure}
\begin{center}
\vspace{1cm}
\epsfxsize=3.0in
\epsfbox[61 199 549 589]{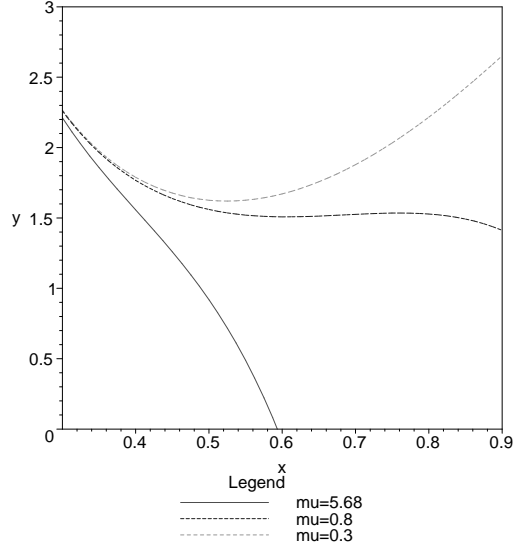}
\end{center}
\caption{ In this figure we  plot the value of the total energy of the QCD ball
per baryon charge (\ref{y},\ref{5b}) with $\sigma\equiv 0$. Three curves corespond 
to the different parameters  $(\frac{4}{\pi})^{2/3}\frac{\mu_c^2}{\sqrt{E_B}}=5.68 ~, 0.8~, 0.3$
modeling  the variation of the bag constant with the baryon density.
The main observation: a minimum corresponding to the equlibrium   does not exist for
the physical parameters (\ref{parameter}) when  $(\frac{4}{\pi})^{2/3}\frac{\mu_c^2}{\sqrt{E_B}}=5.68$
from eq. (\ref{5b}).
Equlibrium appears  when   a typical scale for
 the variation is reduced   by an order of magnitude.}
\label{Fig1}
\end{figure}

a). 
As we already mentioned, in the absence of the axion field, $\sigma\equiv 0$, the problem
was extensively discussed earlier using MIT bag model,\cite{Witten}-\cite{Madsen}.
Our original remark here is: when a variation of the vacuum energy with density is taken
into account, a stable solution {\it disappears} provided that
 a typical QCD scale for
 the vacuum variation (\ref{3b},\ref{3d})  is used.  The physical reason for that behavior
 is quite obvious: a density -dependent vacuum energy is not a sufficiently strong
  squeezer to equilibrate the fermi pressure. A typical scale for
 the variation  should be  reduced (in comparison with what we assumed
in eqs.(\ref{3b},\ref{3d}) ) by an order of magnitude, in order for the solution 
 to reappear.
 Specifically, we checked that the equilibrium is possible for $\sigma\equiv 0$
 if coefficient $5.68$ in (\ref{5b}) describing the vacuum energy variation 
 is replaced by $0.5$.  
We demonstrate this effect in Fig.1 where we display the total energy of QCD ball
per baryon charge (\ref{y}) with $\sigma\equiv 0$ for three different
values of parameter 
$(\frac{4}{\pi})^{2/3}\frac{\mu_c^2}{\sqrt{E_B}}$ describing the effect
of variation of vacuum energy with the baryon density. For a typical choice of
physical  values (\ref{parameter}) the relevant parameter is
 $(\frac{4}{\pi})^{2/3}\frac{\mu_c^2}{\sqrt{E_B}}=5.68$.  In this case
the minimum does not exist which implies that the stability can not be achieved 
as announced above. The minimum starts to reappear only when   a typical scale for
 the variation of density is considerably reduced, see Fig.1 with the curve corresponding
$(\frac{4}{\pi})^{2/3}\frac{\mu_c^2}{\sqrt{E_B}}=0.3$. Such a small value for 
the critical value $\mu_c^2$ does not look appealing from the physics point of view.
 Therefore, we incline to accept that there is no solution for such a
 configuration (strange quark nuggets,\cite{Witten}-\cite{Madsen}) in QCD
 if no external pressure (such as gravity or axion domain wall) is applied.
 It is certainly  not a very new result: special study 
 on stranglets reveals\cite{Alberico} a strong  model dependence of 
the stability of strange quark matter. In particular, the Nambu Jona- Lasinio 
 model does  not support any kind of strangelets\cite{Ratti}.
%
 
b). In general, one expects there existence of a 
 minimal and maximal sizes  
for the QCD balls in the region of stability. The minimal charge $B_{min}$
corresponds to the maximum $\sigma_0^{max}\sim B^{-1/3}_{min}$.
At this point the  
 the stability requirement (\ref{stability}) is marginally satisfied. 
When $B < B_{min}$, $\sigma_0$
  becomes too large such that  
  nucleons can leave the system. 
On the other hand, the maximum possible charge, $B_{max}$,
corresponds to the minimum value of  $\sigma_0^{min}\sim B^{-1/3}_{max}$.
For larger $B$, the baryon density  
(\ref{density}) becomes too low to justify our approach (based on the  quark degrees of freedom).
At lower baryon densities some metastable states may form; they could decay to some heavy elements
which might be of interests for astrophysics. 
However the corresponding study
would  require  an analysis of the system in terms of nuclear degrees of freedom,
which  is beyond  the scope of the present work. When $\sigma_0$ becomes even smaller, 
the problem is essentially equivalent to $\sigma =0$ studied earlier  where
stable solutions are not expected to occur.
\begin{figure}
\begin{center}
\vspace{1cm}
\epsfxsize=3.0in
\epsfbox[61 199 549 589]{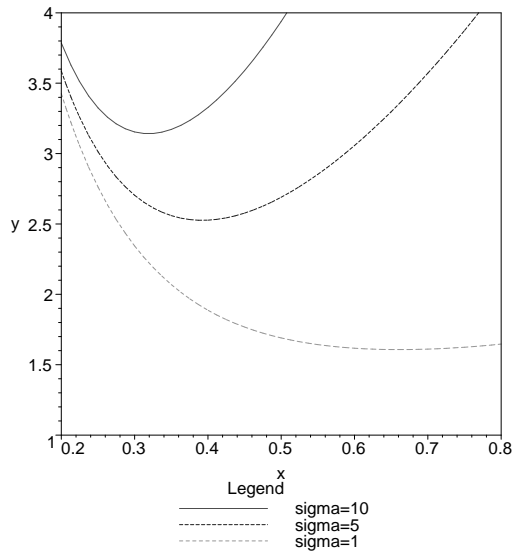}
\end{center}
\caption{ In this figure we plot the value of the total energy of QCD ball
per baryon charge (\ref{y},\ref{5a}) for  $4\pi\sigma_0=1, 5, 10$.
The main observation:  a minimum describing  the equilibrium at $4\pi\sigma_0=1$
 corresponds to the maximum  possible  baryon charge. Solution goes away for smaller $\sigma_0$.
The equilibrium at $4\pi\sigma_0=10$
 corresponds to the minimum   possible baryon charge $B_{min}$  when solution at the equilibrium 
still satisfies the stability requirement (\ref{stability}). At larger  
$\sigma_0 > \sigma_0^{max}$ the quark energy per baryon charge becomes large enough
such that nucleons can leave the system. }
\label{Fig2}
\end{figure}

Numerically, we analyzed two models (\ref{5a}, \ref{5b}) which lead to the similar results.
In particular, for model (\ref{5a}) the maximum possible tension, $4\pi \sigma_0^{max}\simeq 10$
corresponds to the minimum baryon charge $B_{min} $. For  such $\sigma_0$
 the equilibrium is reached at $x_0\simeq 0.32$ 
wnen the  energy per quark $y_{QCD}^{(1)}(x_0)\simeq 2.1$
 hits the upper energy bound of the stability region (\ref{stability1}).
When $4\pi \sigma_0^{max}> 10$, the  energy per quark becomes too high such that nucleon can escape
and the system would decay. In physical units this solution corresponds 
to $B_{min}\simeq 10^{32}$  and  stabilization radius 
$ R_0= x_0\sqrt[3]{B}/\sqrt[4]{E_B}\simeq 10^{11} GeV^{-1}$. Energy per
quark for this configuration $\epsilon_{QCD}^{(1)}=y_{QCD}^{(1)}(x_0)\sqrt[4]{E_B}
\simeq 2.1\sqrt[4]{E_B}\simeq 320 MeV$ is smaller than constituent quark mass, 
as it should be\footnote{We remind that we discuss the QCD part  of energy only;
 the total energy of the configuration which includes the axion part is larger.}.

For the same model, the minimum possible tension when our approach is
justified, $4\pi \sigma_0^{min}\simeq 2$
corresponds to the maximum possible  baryon charge $B_{max}$.
According to the scaling $B\sim \sigma_0^3$, the maximum baryon charge 
$B_{max}=(\frac{\sigma_0^{max}}{\sigma_0^{min}})^3B_{min}\sim 10^{34}$ is 
two orders of magnitude larger than $B_{min}$.
 In this case
the equilibrium is reached at $x_0\simeq 0.52$ when the baryon density (\ref{density}) 
is  already relatively low, and  close to the boundary when the quark based lore can not be 
trusted. 

Our second model (\ref{5b}) gives quantitatively similar results, 
and it is not worthwhile to discuss numerical details here. The most important 
features of the solution for this model remain the same: 
there is a region between $ B_{max} $ and $ B_{min} $
when solutions are stable; at $\sigma=0$ solution does not exist at all
provided that
 a typical QCD scale for
 the vacuum variation (\ref{3b},\ref{3d})  is used. 
 
However, one should take all these numerical estimates very  cautiously
because of a number approximations we have made in eqs. (\ref{1}, \ref{2}).
  Nevertheless, in what follows, mainly for
 the illustrative  purposes, we shall stick with these numerical estimates. 

 The    quark number density $n$ in the 
region  $ B_{min} < B <  B_{max} $
when   our approach is justified is estimated  as
\be
\label{7}
 n \equiv \frac{B}{V}  = \frac{3 E_B^{3/4}}{4\pi x^3}
\simeq (1.5 - 6.5 ) \cdot 3n_0,
\ee
where we used    the
expression (\ref{1}) for the baryon density. As we already mentioned
the expression (\ref{1}) is  formally  valid only for noninteracting quarks, 
but numerically remains a good approximation in a large region of $\mu$.
In eq. (\ref{7}) $3 n_0\simeq 3(108 MeV)^3,$ is the nuclear saturation density normalized with our convention
( $B=1$ for quarks), thus factor $3$ in front of the numerica
value $0.16 (fm)^{-3}\simeq (108 MeV)^3$. It is quite remarkable that the numerical value
for $n$ is in the   region where color superconductivity phase is  likely   to   realize,
and therefore, our treatment of the squeezed fermi system as the quark
dense matter (rather than ordinary nuclear matter)  is  justified a posteriori.

Few remarks are in order regarding eq.(\ref{7}). 
 First of all, the estimates presented above demonstrate that we are in the region
of the phase diagram where CFL phase is likeley to   realize. Therefore, our original
assumption is justified. Secondly,
    for   large $B\geq  B_{max}$
 our treatment of the system  is not valid anymore, and 
 a different type 
of QCD balls with an ordinay nuclear matter (instead of diquark condensate)  in the bulk
may be formed and could be even  stable in some regions of parametrical space.
Though this region of  large $ B\geq B_{max}$ could be an interesting region
from the phenomenological point of view, it shall not be discussed here\footnote{The 
corresponding proper treatment would require the knowledge of the dynamics
of the interacting
nuclear matter, which is not the subject of the present work. 
In principle such nuclear matter
could be also stable.}.
  However, even in this case when the QCD balls
made of nuclear matter, rather than quark dense matter,
we still expect that there   should exist
 a maximum size    above which the stability is not possible.
This follows from our analysis that stability can not be achieved without the external pressure
$P_{\sigma}$  due to  the axion domain $P_{\sigma}\sim 2\sigma/ R$ 
  which vanishes at very large $R$. 
 
Another factor which also constraints the size of the balls is related to the suppression
 of  large size closed axion domain walls during the formation stage. 
 It is clear that the formation of the large size closed domain walls is
 suppressed according to the
 Kibble-Zurek  mechanism\cite{Kibble},\cite{Zurek}; however 
 an explicit estimation for this effect is still missing.

As we mentioned in the Introduction, we do not address the problem of formation of QCD balls
in this letter, it will be a subject of a different work.  However we would like to
mention some relevant elements of a possible scenario of how QCD-balls, in principle,
can be formed after the QCD phase transition, at a temperature of
order 150 MeV which is much higher than the critical
temperature for quark pairing   estimated to be $\sim 0.6\Delta$.  
The main point is this: the axion domain wall with the QCD-scale substructure
as discussed in\cite{FZ} is very selective with respect to the momentum of the particles;
it is almost transparent for light $\pi$ mesons with 
large momentum  $k \geq m_{\pi}$ such that the  transmission coefficient is close to one.
Therefore, the  highly energetic pions  can easily penetrate through the domain wall and leave
 the system. 

At the same time, the  transmission coefficient is close to zero for slow-moving
particles such as baryons with $k \leq m_{\pi}$. Eventually, this  ``selective"  feature
 of the domain wall may cool down the system considerably. Due to the 
domain wall pressure it may reach the critical 
density when it undergoes a phase transition to a color  superconducting phase
with the ground state  being the  quark condensate. 
At this point we assume that the baryon number trapped in the bulk is sufficiently large.
If $B\gg B_{max}$, the  quarks will leave the system by forming
nucleons until the upper limit $B_{max}$ is achieved. At this point the 
   energy per unit baryon charge 
$\epsilon_{QCD}=\sqrt[4]{E_B}y_{QCD}(x_0) < \frac{m_N}{3}$,  
 becomes  sufficiently small such that     quarks  can not leave the system.
Some specific calculations are required before any statements regarding a possibility
to form the  QCD balls   can be made. We do not see any fundamental obstacles which would
prevent the  formation of such objects. Terefore, 
at this moment we simply assume
that this probability does not vanish. 
 
\subsection{QCD- balls versus Q-balls}
In this subsection we would like to mention a striking  resemblance
of the QCD-balls (which is the subject of this letter) and
non-topological solitons\cite{Lee}, as well as  Q-balls\cite{qball} 
which is a special case of a nontopological soliton configuration
associated with some conserved global $Q$ charge.  
Both cases, QCD balls and Q-balls demonstrate a similar behavior
for a 
soliton mass as function of $Q$.  Namely, QCD balls as well as Q-balls
  may become very  stable configurations for relatively large $Q$ charge. Therefore, an effective
 scalar field theory
with some specific constraint on the effective potential (when
$Q$ ball solution exists) is realized for QCD in high density regime by 
formation of the diquark scalar condensate which plays the role
of the effective scalar field.
The big difference, of course, that underlying theory for QCD-balls is well known, it is QCD
 with no free parameters. This is  in   contrast with the theory of   Q-balls
when the underlying theory is not known. 
Formal similarity becomes even more striking if  one takes into account that the ground state of the
CFL phase in QCD  is determined by the diquark condensate with the following  time 
dependence $\sim e^{i2\mu t}$,
\be
\label{diquark}
    \langle \Psi_{La}^{i\alpha}\Psi_{Lb}^{j\beta}\rangle^* \sim
    \langle \Psi_{Ra}^{i\alpha}\Psi_{Rb}^{j\beta}\rangle^* 
    &\sim (e^{i2\mu t})\cdot\epsilon^{ij}\epsilon^{\alpha\beta c}\epsilon_{abc}
    ~~,  
\ee
with $\Psi$ being the original QCD quark fields, and $\mu$ being the chemical potential of the system,
see formula (40) from ref. \cite{FZ2}.
As is known, such time-dependent phase is the starting point in construction of the 
Q balls\cite{qball}.
 In the expression (\ref{diquark}) we  explicitly show the structure
for the diquark condensate corresponding to CFL (color-flavor locking) phase\cite{cs_r}
with   ($\alpha$, $\beta$, etc.) to be flavor,
($a$, $b$, etc.) color  and  ($i$, $j$, etc.) spinor indices correspondingly.
Of course, there are many differences in phenomenology between Q balls\cite{qball}  and QCD-balls. 
For example, in CFL phase the baryon symmetry is spontaneously broken, and corresponding Goldstone
massless boson
carries the baryon charge. However, the evaporation of this massless particle into hadronic phase
from the surface of the QCD-ball is not possible, because hadronic phase does not support such excitation.
This is in contrast with phenomenology of Q-balls, where the theory is formulated in terms
of one and the same scalar $\phi$ field, such that evaporation of $\phi$ particles 
from the surface of the Q-ball is possible if some conditions are met.
In spite of many differences, the  analogy with Q-balls is quite useful and can be used for analysis
of different experimental bounds  on QCD-balls, which is the subject of the next section.

\section{Experimental bounds on masses and fluxes of QCD-balls}
In this section we adopt the results of paper\cite{qball_exp} to constraint 
the free parameter (charge $B$) of the QCD-balls.
In the paper \cite{qball_exp} the authors re-analyzed  the results of various experiments, 
originally not designed for the Q-ball searches, but nevertheless these experimental results 
were successfully used 
in \cite{qball_exp} to bound  different properties of the Q-balls. We actually repeat this analysis 
for a specific type of the QCD-balls when original quarks are in the CFL (color-flavor locking)
phase\cite{cs_r}.

As we mentioned earlier, at sufficiently large baryon density, the color superconductivity 
phenomenon takes place.
However, there are many different 
phases (as a function of parameters like $m_s$, number of light flavors, etc.) associated with
color superconductivity. In particular, for 3 degenerate flavors of light quarks, the CFL phase
with nonzero value for the diquark condensate (\ref{diquark}) is realized. Due to the fact that 
equal numbers of $u, d, s$ quarks
condensed in the system, the electric charge of the ground state is zero, i.e.
 no electrons required to neutralize
the system. This is quite important feature for the phenomenology of the QCD-balls
 we about to discuss.
Nature is less symmetric, and other CS phases could be realized. In particular, 
for relatively large $m_s$, along with diquark condensate, the $K$ condensate
 may also be formed\cite{Thomas}.
In the limit of  very large $m_s$, QCD becomes effectively a theory with two light quarks.
In this case, the Cooper pairs are $ud-du$ flavor singlets. This phase,   the so-called 2SC
(2 flavor super-conductor ) phase is a phase with non-zero electric charge. Electrons
 neutralize the system,  
however, all properties, such as interaction cross sections, the rate of energy loss of QCD balls  
 in matter, 
 are very different for QCD-balls with quarks in CFL or 2SC phase.
In what follows, to avoid many complications,  we limit ourself with analysis of QCD balls 
where quarks are in the most symmetric CFL phase, in which case the QCD-ball has zero electric charge.

We assume, in analogy with\cite{qball_exp}, that a typical cross section of a neutral QCD-ball
  with matter is determined by their geometrical size, $\pi R_0^2$.
In this case,   the only information we need to constraint the QCD-ball parameters,
is its size   and mass. We also assume that the QCD-balls is  the main contributor 
toward the dark matter in the Galaxy. Their flux $F$ then should satisfy
\be
\label{dark}
F < F_{DM}\sim \frac{\rho_{DM} v}{4\pi M_B} \sim 
7.2\cdot 10^5 \frac{ GeV}{M_B} cm^{-2}sec^{-1}sr^{-1},
\ee
where $\rho_{DM}$ is the energy density of the dark matter in the Galaxy, 
$\rho_{DM}\simeq \frac{0.3 GeV}{cm^3}$, and $v\sim 3\cdot 10^{-3}c$ is
 the Virial velocity of the QCD-ball.
We identify $M_B$ in the expression (\ref{dark}) 
with the total energy $E$  of the QCD ball at rest with given baryon  charge $B$.
The {\it  Gyrlyanda} experiments at Lake Baikal 
reported that the flux of neutral soliton-like objects has the bound
\cite{Baikal}
\be
\label{8}
F~ <~ 3.9\cdot 10 ^{-16} cm^{-2}sec^{-1}sr^{-1},
\ee
which translates to the following 
lower limit of the neutral QCD-ball mass $M_B$ and baryon charge $B$,
\be
\label{9}
M_B^{exp} ~ > ~ 2\cdot 10^{21}~~ GeV ,~~~~~~~~~~~~~~~~~~~~~~~~~~~~ \\
B^{exp} \simeq  (\frac{M_B}{\sigma^{1/3}})^{9/8} [\frac{3}{2}(8\pi c^2)^{\frac{1}{3}}]^{-9/8}
~ >~ 1.6\cdot 10^{20}. \nonumber
\ee
Similar constraints follow from the analysis of the {\it Baksan} experiment\cite{Baksan}
and analysis \cite{qball_exp}
of the Kamiokande Cherenkov detector\cite{Kam}, and we do not explicitly quote these results.
These experimental bounds are well below the critical line of the  stability
of the QCD-balls.    

 \section{Discussions and Future directions}
Complete theory of formation of the QCD-balls is   still lacking. Only such a theory would predict whether
QCD-balls can be formed in sufficient number to become the dark matter.
Such a  theory of formation of the QCD balls would answer on questions like this: 1. What is the 
probability to form  a closed axion domain wall with size $\xi$ during the QCD phase transition?
 2. How many quarks are trapped inside the domain wall at the first instant? 3. How many quarks
will leave the system and how many of them will stay inside the system while the bubble is shrinking? 4. 
What is the  dependence of  relevant parameters such as: size $\xi(t)$, baryon number density $n(t)$
and internal temperature $T(t)$ as function of time? 5. Do these parameters fall into appropriate region
of the QCD phase diagram where the color superconductivity takes place? 6. What is the final density
distribution of the QCD-balls as a function of their size $R$ after the formation
period is complete?  7.Will the QCD balls survive the evaporation and boiling even if they formed?
Clearly, we do not have answers on these, and many other important questions at the moment.
All these interesting, but difficult questions   are obviously beyond the scope of the present work,
and shall not be discussed here. However, we want to make a short comment on issue 7
which was an important element in many previous studies.
 
The question  on evaporation of quark nuggets was discussed earlier, see 
original papers \cite{Alcock}-\cite{SK} and recent review \cite{Madsen}. The first study
of this question is due to Alcock and Farhi\cite{Alcock} who argued that only
very large nuggets with $B \geq 10^{52}$ could survive the evaporation. This result
would essentially eliminate the possibility of any quark nuggets surviving till the present
epoch. However, Madsen et al.\cite{Madsen1} then point out that few important effects
can considerably reduce the original estimation given in ref.\cite{Alcock}.
The first important effect   is related to the 
deficiency of $u$ and $d$ quarks (in contrast with  $s$ quark)
in the surface area. This leads to the suppression of the 
evaporation rate such that $B \geq 10^{46}$ can be stable against evaporation\cite{Madsen1}. 
In this calculation the penetrability of the phase boundary was assumed to be near $100\%$.
This assumption was questioned in \cite{Bhattacharjee} and \cite{SK} where it was demonstrated
that nuggets with $B \geq 10^{43}$ \cite{Bhattacharjee}($B \geq 10^{39}$ according to ref. \cite{SK})
 could survive the evaporation even if the first  effect ( described above 
and which led to $10^{-6}$ suppression, see \cite{Madsen1}
for details) is neglected.  As discussed in \cite{Alcock}-\cite{SK} the limit on $B$ may be further
reduced by reabsorption. All these effects taken together suggest that  
nuggets with $B \geq 10^{30}$ are not ruled out and can survive the evaporation\cite{Madsen}.

Our original remark here is: along with the suppression effects discuseed
above, we have two additional effects which may further reduce the evaporation rate.

 Indeed, the core  of the axion domain wall  
as discussed in\cite{FZ},\cite{SG} has a QCD sub-structure with
a typical scale $\geq 1 GeV$. 
It is quite obvious that this sub-structure certainly reduces  the penetrability
of particles from inside to outside,
and therefore, it suppresses the evaporation rate. Also, the baryon charge 
in superconducting phase is in the form of the diquark condensate 
rather than in form of free quarks discussed in the
previous analysis\cite{Alcock}-\cite{SK}. This fact may also  considerably reduce the evaporation rate
because it requires the  breaking of
 the Cooper pair before the evaporation becomes possible.
This effect certinly increases the effective binding energy and decreases the evaporation rate.
It is difficult to make a precise estimate of these effects at the moment,
due to the many compications discussed earlier\cite{Madsen}
 as well as many additional difficulties mentioned above.
 However, we believe,  it is fair to say that  
the QCD balls with $B\geq 10^{32}$ as discussed in the previous section,
can safely survive the evaporation, 
and therefore, the possibility seems worth exploring.

Now,    we  wish to estimate the absolute value 
for the dark matter number density $n_{DM}$ assuming   
   that the nonbaryonic dark matter is 
actually the QCD balls.
In this case, at 
the QCD phase transition at $T\sim T_c$ soon after the QCD balls are formed,
$n_{DM}$ can be estimated as follows,
\be
\label{r4}
n_{DM}\sim 5\cdot 10^{-9}\frac{2\pi^2}{45}g_*T_c^3 \frac{m_N}{M_B},
\ee
where we used the known magnitudes for the baryon to photon ratio,
 $n_B/n_{\gamma}\simeq 5\cdot 10^{-10}$,  and
the dark matter to baryon ratio, $\Omega_{DM}/\Omega_B\simeq 10$.
Numerically,  for the baryon  charge $B\sim 10^{32}$  and effective
massless degrees of freedom, $g_*\simeq 10$
the estimation (\ref{r4}) leads to 
\be
\label{r5}
r T_c\equiv n_{DM}^{-1/3}T_c \simeq 3.5\cdot 10^{13},~~ r\sim  10 ~cm,
\ee
where $r$ has the  physical meaning of an average distance between QCD-balls after they formed.
As expected, average distance $r$ is much smaller than the horizon radius $R_H^{QCD}$
at the QCD phase transition, $r\sim 10^{-5}R_H^{QCD}$.
It is quite remarkable that $r$ is much larger than the size of the
 QCD-ball, see   eq.(\ref{2}), 
such that QCD-balls become well separated soon after they formed.
Besides that we expect that the QCD ball size should be related, through dynamics,
 to the correlation length $\xi\sim m_a^{-1}$ of the original axion field.
We also expect that the spatial extend of a typical
closed wall at the instant of formation has the same order of magnitude 
$\xi ~$ \cite{Kibble, Zurek}.  Initial size of a closed wall $\sim\xi$
eventually (after some shrinking as a result of tension, and after
some expansion as a result of evolution of the Universe) 
determines  the size of the QCD-balls. However, the dynamics of this transition
is quite complicated, and we are not able to derive a relation between initial domain wall size 
distribution and QCD-ball size distribution at the later stage.
  Close numerical values for the QCD ball size
  and $\xi\sim m_a^{-1}$ also suggest that these parameters are related somehow.
 Therefore, it is at least possible,
that the decay of the axion domain wall network may result in formation of the QCD-balls
with their nice properties discussed in this work.

To conclude: we advocate the idea that the QCD-balls could be 
a viable cold dark matter candidate which
 is formed from the ordinary quarks
during the QCD phase transition when  the  axion domain walls
form.    As we argued 
the system in the bulk     may reach the critical 
density when it undergoes a phase transition to a color  superconducting phase
in which case the new state of matter
representing the diquark condensate with a 
large baryon number $B$
  becomes  a stable soliton-like  configuration. 
The scenario is no doubt lead to important consequences for cosmology and astrophysics,
which are not  explored yet.
In particular,  some unexplained events, such as Centauro events, or  even the Tunguska-like
events (when no fragments or chemical traces have ever been recovered),
can be related to the very dense QCD balls.
The recent detection\cite{Teplitz} of two seismic events with epilinear
(in contrast with a typical epicentral ) sources may also be related to the 
very dense QCD balls. Also, the ``missing" baryons in Galaxy Clusters\cite{Ettori}
may also be related to the QCD balls. Finally, the cuspy halo problem in dwarf
galaxies might be related to the unstable cold dark matter \cite{Cen},
which, again, could be  related to the    QCD balls discussed in this work.
Indeed, as we mentioned in the Introduction, if the QCD ball size 
exceeds the critical value, it becomes metastable (rather than
stable) configuration. The  life time of these metastable QCD balls
could be very large. Therefore, they could serve
as decaying dark matter particles suggested in\cite{Cen}.

 Therefore, the ``exotic", dense color superconducting phase in QCD, might be much  
more common state of matter in the Universe  than the ``normal" hadronic phase we know.
More than that: one can present some arguments\cite{OZ} to support the idea that the observed
in nature asymmetry between baryons and antibaryons may also be originated from the 
same physics during the QCD phase transition.   In this case the antimatter is hidden 
inside of the anti- QCD balls in the form of the diquark condensate similar 
to the QCD ball case. One could naively think that such
a scenario is in contradiction with observations on absence of antimatter around us.
However, such a  conclusion would be very premature one due to 
  the specific  interaction features of the matter in hadronic phase  with the matter in  color 
superconducting phase.
Namely, 
if the energy of the quark  which hits the anti-QCD ball is smaller than the 
superconducting gap $\Delta$, the quark can not penetrate into the bulk,
break the Cooper pair and excite
the internal degrees of freedom of the system.   
Rather it will be reflected \cite{OZ}.
A similar property is commonly  known 
as the ``Andreev Reflection" in the literature on 
conventional superconductivity.  
Therefore, at low energies,
the anti-QCD balls   behave as QCD balls
with respect to the interaction with environment,
 and there is no contradiction with known constraints on  
such kind of anti matter in our Universe.

In this case, without fine tuning of parameters, 
one can easily   understand the  relation between $\Omega_{DM}\sim \Omega_B $ 
which both originated at the same instant. As is known
 this  ratio  is very difficult to understand 
if these quantities do not have the same origin.  

In conclusion, qualitative as our arguments are, they suggest that
 the dark matter   could be  originated at the QCD scale.

\section*{Acknowledgments}
I am thankful to Robert Brandenberger,    Michael Forbes,   
Sasha Kusenko, Jes Madsen, Shmuel Nussinov, Krishna Rajagopal, 
Edward Shuryak, Pierre Sikivie, Paul Steinhardt 
and Frank Wilczek  for useful discussions, comments and  remarks.
This work was supported in art by the National Science and Engineering
Research Council of Canada. 
\section*{Appendix}
The main goal of this Appendix
is to argue that the probability of the  absorption of axion
by emitted quark is extremely small. In this case   our criteria 
of stability of the QCD balls, see eq.(\ref{stability}),
which neglects the axion domain wall energy contribution, is justified.
Such a treatment of the problem essentially implies that
we impose a weaker condition of metastability (rather than
a stronger condition of the absolute
stability) on the QCD balls. In different words, we assume that the energy can not
be   transformed from the axion domain wall to the   quark
which is about to leave the system. 
In what follows we make
some estimates which support this assumption.
%
Indeed, as we shall see in a moment
this probability is exceedingly small due to the very small nucleon-axion coupling constant,
$ \frac{m_q}{f_a}\sim (10^{-13}-10^{-15})$. 

We start from  the estimation of the probability to
 absorb the axion from the axion domain wall background field
$a(z)$ by an elementary excitation $|\psi_{in} >$ with mass $\Delta$ proportional to the gap.
 This elementary excitation carries the unit baryon chage in superconducting phase. We assume that the final state is   represented by the  wave function of 
constituent quark $< \psi_{out} |$  with mass $\sim m_N/3$  ( hadronic phase).
We take a simpe expression for the  axion-quark interaction to be $\frac{m_q}{f_a}\psi^{\dagger}_{out}a(z)\psi_{in}$,
where $m_q$ is the current mass quark of order few $MeV$, and $\frac{a(z)}{f_a}\sim 1$
is the axion domain wall background field. The precise expression for the domain wall
profile function $a(z)$ is known, however, in our estimate we shall use
a simple expression $a(z)\sim f_a e^{-m_az}$ in order to emphasize that
 the magnitude of the axion field vanishes at infinity and the 
typical scale where axion field varies is $m_a^{-1}$.
We also assume that the quark has a trajectory $z=vt$ with velocity $v$ close to the speed of light.
In this case the time-dependent interaction takes the form $\sim m_qe^{-m_at}$, and the probability
for the transition 
can be estimated from the dimensional arguments as follows, 
\be
\label{absorption}
W\sim |m_q\int_0^{\infty}dte^{i\omega t-m_at}+h.c.|^2\sim |\frac{m_qm_a}{m_a^2+\omega^2}|^2,
\ee
where $\omega$ is the energy difference between $|\psi_{in} >$ and $ |\psi_{out} > $ states.
It is impotant to note that, typically $\omega\simeq (100-200) MeV$ is large,  and therefore, the probability
(\ref{absorption}) is very small. We neglected many factors in estimate (\ref{absorption}).
In particular, we neglected the  momentum dependence of $|\psi_{in} >$ and $ |\psi_{out} > $ states;
the mismatch between these momenta would bring an additional suppression to (\ref{absorption}), 
and  we ignore 
this effect at the moment. It is easy to understand the source of the suppression in 
eq.(\ref{absorption}):
the probability for a considerable  excitation $\sim \omega$ of  the system by   a smooth
field with a typical correlation scale $m_a^{-1}$ is very small. 
 
 In order to derive
a total number of events of absorption $W_{tot}$
one should multiply the expression(\ref{absorption})
 by an additional factor  
describing  a total number of elementary quark excitations 
close to the surface of the system such that
they can leave the system without re-scattering. This requirement (to be close to the
surface of the QCD ball) is important because the distance from the surface
should not exceed the mean free path.
Otherwise, the quark even if it absorbs the axion,  would not be 
able to leave the system.    Assuming the thermodynamical
equilibrium  at temperature $T$
soon after the formation of the QCD balls, we can  estimate this factor  
 as follows $\frac{2\pi^2}{45}g_*T^3\exp(-\frac{\Delta}{T})4\pi R^2\xi, $
where $\xi $ is the mean free path which we estimate to be $1/T$.
Our final expression for the total probability of absorption of the axion
  (while the temperature is   of order $T$) is estimated to be
\be
\label{absorption1}
W_{tot}\sim |\frac{m_qm_a}{m_a^2+\omega^2}|^2
\frac{2\pi^2}{45}g_*T^24\pi R^2\exp(-\frac{\Delta}{T}), 
\ee
where we neglected many additional suppression factors, such as  factor $1/6$ 
describing the probability for the quark to  move in the   direction 
pointing   off the center of the QCD ball.
Numerically, even  if we neglect the factor $\exp(-\frac{\Delta}{T})$
in eq.(\ref{absorption1}), the probability
is already quite small, $W_{tot} < 10^{-3}$ for 
the typical values of $g_*\sim 10~~, T\sim 0.6\Delta $.  When temperature becomes 
considerably  smaller
than  $T$, the probability
of absorption diminishes   due to the small number of excitations, 
$\sim \exp(-\frac{\Delta}{T})$. This late epoch of evolution can be ignored.
 Also, one should keep in mind that the 
quark excitations are not supported in the hadronic   phase due to the
confinement. Therefore, one should have  three quarks (or quark and diquark pair 
from the condensate) to be  organized in a color singlet state
such that it  can propagate in the hadronic phase.
It   definitely  gives an additional suppression which we even did not try to estimate:
the suppression factor (\ref{absorption1}) is already sufficiently strong for our purposes.
Therefore,  our treatment of the problem when we use a 
weaker condition of metastability, see eq. (\ref{stability}),
 rather than
a stronger condition of the absolute
stability, is justified.

\end{document}